\documentclass[11pt, executivepaper]{article}
\usepackage[utf8]{inputenc}
\usepackage[T1]{fontenc}
\usepackage{natbib}
\usepackage{amsmath}
\usepackage{mathtools}
\usepackage{xcolor}
\usepackage{amsfonts}
\usepackage{graphicx}
\usepackage{enumitem}
\usepackage{geometry}
\usepackage{hyperref}
\hypersetup{colorlinks= true, allcolors=blue}
\setcitestyle{aysep={}}
\begin{document}

\title{\textbf{The Bundle Theory Approach to Relational Quantum Mechanics}}

\author{Andrea Oldofredi\thanks{Contact Information: Universit\'e de Lausanne, Section de Philosophie, 1015 Lausanne, Switzerland. E-mail: Andrea.Oldofredi@unil.ch}}

\maketitle

\begin{abstract}
The present essay provides a new metaphysical interpretation of Relational Quantum Mechanics (RQM) in terms of mereological bundle theory. The essential idea is to claim that a physical system in RQM can be defined as a mereological fusion of properties whose values may vary for different observers.\ Abandoning the Aristotelian tradition centered on the notion of substance, I claim that RQM embraces an ontology of properties that finds its roots in the heritage of David Hume.\ To this regard, defining what kind of concrete physical objects populate the world according to RQM, I argue that this theoretical framework can be made compatible with (i) a property-oriented ontology, in which the notion of object can be easily defined, and (ii) moderate structural realism, a philosophical position where relations and relata are both fundamental. Finally, I conclude that under this reading relational quantum mechanics should be included among the full-fledged realist interpretations of quantum theory. 
\vspace{4mm}

\end{abstract}
\vspace{5mm}

\begin{center}
\emph{Accepted for publication in Foundations of Physics}
\end{center}

\clearpage 
\tableofcontents
\vspace{10mm}

\begin{flushright}
\emph{The world divides into facts.}

L. Wittgenstein, \emph{Tractatus Logico-Philosophicus}, Proposition 1.2
\end{flushright}
\vspace{5mm}

\section{Introduction}
\label{Intro}

Relational Quantum Mechanics (RQM) is an interpretation of quantum theory proposed by Carlo Rovelli in several publications (cf.\ \cite{Rovelli:1996, Rovelli:2016, Rovelli:2018}, \cite{Laudisa:2019b}). According to this interpretative framework, Quantum Mechanics (QM) concerns to the observable properties of physical systems \emph{relative} to specific observers. In this context, indeed, quantum systems can be described differently by distinct observers, and these diverse representations are not contradictory. Thus, the relational character of RQM entails that the notion of an absolute, unique reality is abolished.

Such a theory is motivated by Rovelli's pioneering work in quantum gravity\footnote{For details about Rovelli's research on quantum gravity the reader may refer to \cite{Rovelli:2004} and \cite{Rovelli:2014}.}, as explicitly stated in \cite{Rovelli:2014} and more recently in \cite{Rovelli:2018}. In loop quantum gravity, in fact, one does not rely on a background spacetime as a container, or an arena where to ``locate things''. Rather, the fundamental items of this framework somehow create it---or better, they reproduce its salient features---providing a \emph{relational} perspective about this notion. Moreover, Rovelli argues that also Einstein's theory of general relativity endorses a relational view about spacetime, since the localization of dynamical items---i.e.\ spacetime regions, the gravitational field, \emph{etc.}---is not determined with respect to a fixed background. Similarly, RQM underlines the \emph{relational} character of quantum theory:
\begin{quote}
[t]he theory yields probability amplitudes for processes, where a process is what happens between interactions. Thus quantum theory describes the universe in terms of the way systems affect one another. States are descriptions of ways a system can affect another system. Quantum mechanics is therefore based on \emph{relations} between systems, where the relation is instantiated by a physical interaction (\cite{Rovelli:2014}, p.\ 54).
\end{quote}
\noindent Referring to this, Rovelli more explicitly states that
\begin{quote}
a quantum mechanical description of a certain system (state and/or values of physical quantities) cannot be taken as an ``absolute'' (observer-independent) description of reality, but rather as a formalization, or codification, of properties of a system \emph{relative} to a given observer.\ Quantum mechanics can therefore be viewed as a theory about the states of systems and values of physical quantities relative to other systems. [...] Therefore, I maintain that in quantum mechanics, ``state'' as well as ``values of a variable''---or ``outcome of a measurement''--- are relational notions in the same sense in which velocity is relational in classical mechanics'' (\cite{Rovelli:1996}, pp.\ 1648-1649).
\end{quote}

Remarkably, the relational character of general relativity can be made compatible with that of RQM in virtue of the notion of \emph{locality}---a central concept in Einstein's theories. Indeed, \cite{Rovelli:2018} claims that if we identify the quantum mechanical notion of ``physical system'' with the general relativistic concept of ``spacetime region'', then the notion of ``interaction'' between systems in QM becomes specular to that of ``adjacency'' between spacetime regions: locality assures that interaction requires physical adjacency. As a consequence, 
\begin{quote}
quantum states are associated to three dimensional surfaces bounding spacetime regions and quantum mechanical transition amplitudes are associated to ``processes'' identified with the spacetime regions themselves. In other words, variables actualise at three dimensional boundaries, with respect to (arbitrary) spacetime partitions. The theory can then be used locally, without necessarily assuming anything about the global aspects of the universe (\cite{Rovelli:2018}, pp.\ 10-11). 
\end{quote}
This is a perfect exemplification of what has been stated above concerning the rejection of an absolute, global characterization of reality in RQM.\footnote{Analyzing Rovelli's essays on RQM, it is immediately clear that Einstein's relativity theories and their axioms played a crucial role in order to develop the relational interpretation of QM, as evident in \cite{Rovelli:1996}. In the present essay, however, I will not discuss the notion of locality in the context of relational quantum mechanics; the interested reader may refer to \cite{Rovelli:2019} and \cite{Pienaar:2019}.}

Another motivation for a relational interpretation of QM lies in the fact that relational quantities are ubiquitous in physics, from classical mechanics and electromagnetism to quantum field theory and cosmology. For instance, the classical notion of velocity makes sense only relative to a given reference frame, potentials of a single point in electromagnetism do not have meaning \emph{per se}, but they acquire physical significance only if another point is taken into consideration as a reference. Similarly, the Unruh effect in quantum field theory affirms that, from the point of view of an accelerating observer, the vacuum of an inertial observer will contain a gas of particles at a temperature which is proportional to the acceleration, so that these two observers will find a different number of particles. Another example is the passage of time which is different from an observer on Earth and an observer located on the edge of a black hole. We also can experience relational phenomena in our everyday life by simply looking at the stars: when observing the light emitted by a star situated very far from us, we are looking at events happened in the far past. Reality, then, is very different from the perspective of the observed star and from ours. 

The main novelty of Rovelli's approach is to take seriously into account the relational character of physics, applying it to the formalism of standard quantum mechanics taken at face value. This strategy entails remarkable consequences not only for the \emph{Weltanschauung} that RQM provides, but also for the interpretation of the central notion of QM, the wave function. On the one hand, RQM aims at providing a realistic picture of what happens in spacetime starting from the principle that physical systems have properties whose values are observer-dependent---i.e.\  such values may differ from the perspective of two distinct observers. On the other hand, the quantum mechanical wave function $\psi$ loses its central role as observer-independent representation of quantum states. In this respect, relational quantum mechanics is conceptually close to Heisenberg's matrix mechanics, where (i) the ontologically relevant items are observables whose values evolve in time, and (ii) quantum jumps refer to the updates of these values upon interactions. Nevertheless, although Heisenberg's views about quantum theory oscillated from positivism to a form of realism over the years\footnote{The reader may refer to \cite{Jaeger:2009}, Chapter 3, for details about Heisenberg's views on quantum theory.}, Rovelli always emphasized the realist commitment of his interpretation. 

Quite ironically, however, relational quantum mechanics is seldom considered as a realist interpretation of quantum mechanics. Looking at the literature concerning the philosophical foundations of non-relativistic QM, indeed, the proposals usually considered as realistic approaches to quantum physics are Bohmian mechanics (cf.\ \cite{Bohm:1952aa}), the spontaneous collapse theories (cf.\ \cite{Bassi:2003}), the many world interpretation (cf.\ \cite{Wallace:2012aa}), the family of modal interpretations (cf.\ \cite{Lombardi:2017}), \emph{etc.}, but not RQM. Contrary to this attitude, the main aim of the present essay is twofold: (i) to provide Rovelli's theory with a new ontological interpretation, and (ii) to argue that under this novel reading it can be considered a full-fledged realist interpretation of QM.

The ontology of relational quantum mechanics is usually defined in terms of ``events'' or ``facts'', recalling the well-known first proposition of Wittgenstein's Tractatus, where it is stated that the world is the collection of facts, and not of things (cf.\ \cite{Wittgenstein:1921}).\footnote{There is evidence that Wittgenstein's work influenced some reflections on RQM since the Tractatus is explicitly quoted in \cite{Smerlak:2007}.} The notions of events and facts refer in RQM to interactions among physical systems (cf.\ \cite{Smerlak:2007}, \cite{Laudisa:2019b}).\ Nonetheless, one may further clarify the ontology of this theory by asking what physical systems are, given that they interact in spacetime. In the present paper, my contribution is to provide an answer to this question using the tools of analytic metaphysics, defining systems as a mereological bundle of properties which not only can vary in time, but also are observer-dependent. In this manner, the event ontology of RQM can be presented as an ontology of interactions among systems, where the latter are now unambiguously defined. Consequently, I will point out also that RQM does not need to be framed in the Aristotelian tradition, where the notion of substance is primary, but it can be understood in terms of the property realism initiated by David Hume in his \emph{Treatise on the Human Nature} (cf.\ \cite{Hume:2007})---where a devastating criticism to the concept of substance is given---and currently evolved in different versions of bundle theories. Finally, since a metaphysically satisfying definition of physical systems can be given in RQM, I argue that this theory is compatible with moderate structural realism, a position where relations and relata are both fundamental.
\vspace{2mm}

The paper is organized as follows: in Section \ref{RQM} an overview of relational quantum mechanics and its assumptions is given (readers familiar with this framework can skip this section). Mereological Bundle Theory (MBT) is introduced in Section \ref{bundle} and it will be applied to relational quantum mechanics in Section \ref{RQMBTA} in order to provide a new metaphysical interpretation of physical systems in this framework.\ Moreover, in this section it is  argued that Rovelli's theory is compatible with moderate structural realism. Finally, Section \ref{conc} concludes the essay.

\section{A Brief Overview of Relational Quantum Mechanics}
\label{RQM}

It is well-known that in RQM the notion of absolute, observer-independent state of a system withers away together with an observer-independent attribution to values of physical magnitudes, in favor of a relational view of quantum states.\footnote{This presentation of RQM follows closely \cite{Rovelli:1996}.} Indeed, according to relational quantum mechanics the state of a system is meaningfully defined only with respect to another system, which plays the role of an external observer. More importantly, RQM rejects the idea of an absolute, observer-independent reality. Although this is a radical departure from the common sense intuition of the world we live in, Rovelli's relational conception of reality does not lead to subjectivism.\footnote{Interestingly, \cite{Candiotto:2017} convincingly argued that RQM does not entail any form of ontological relativism.} To this regard, the word `observer' refers in RQM to \emph{every} physical object having a particular definite state of motion; in this framework an electron, an air molecule, as well as a Geiger counter or a human experimenter can be observers (cf.\ footnote 9 below).\footnote{To this regard, it is also worth noting that RQM describes \emph{every system} quantum mechanically, regardless if it is microscopic or macroscopic, avoiding by construction the cogent problem of ``where to put the cut'' between the quantum and the classical regime. More on this below.} Hence, this notion is not necessarily referring to conscious beings, which do not have any power to create reality---which exists \emph{per se}---contrary to other approaches to quantum theory. 

According to Rovelli, such a relativization of states and values of physical quantities has a strong empirical basis, which ``derives from the observation that the experimental evidence at the basis of quantum mechanics forces us to accept that distinct observers give different descriptions of the same events'' (\cite{Rovelli:1996}, p.\ 1638). Interestingly, another motivation to rebut the notion of absolute reality comes from a serious consideration of the methodological evolution of Einstein's theory of special relativity. Indeed, Rovelli notes that the difficulty physicists had to interpret correctly Lorentz transformations were due to the concept of \emph{absolute simultaneity}. Removing this notion, Einstein was able to endow such transformations with a physical \emph{interpretation}. Similarly, in relational quantum mechanics it is claimed that the conceptual and technical conundra of standard quantum theory are due to the presence of the physically incorrect notion of \emph{absolute state} of a system. Taking the quantum formalism at face value, and following Einstein's steps eliminating the notion of ``absolute state'', the main aim of RQM is to provide it with a new interpretation in order to understand what this theory tells us about the world. Furthermore, the project of RQM is to re-derive the formalism of quantum theory starting from simple, physically meaningful principles which are experimentally motivated---retracing again the example of Einstein's derivation of Lorentz's transformations from the simple postulates of special relativity.\footnote{For space reasons the details of such a derivation will not be given here; the interested reader may refer to \cite{Rovelli:1996}, Section 3.}

That being said, it is useful to recall what principles of standard quantum mechanics carry over intact in RQM, in order to delineate its formal structure:
\begin{itemize}
\item \textbf{Eigenvalue-Eigenstate Link:} The properties of physical systems in RQM are represented by self-adjoint operators, i.e.\ the usual quantum observables. More precisely, given a quantum observable $A$, a quantum system $s$ possess a sharp value $a$ for the quantity $A$ if and only if $s$ is in an eigenstate of $A$ that correpsonds to $a$;
\item \textbf{The Schr\"odinger Equation (SE):}
\begin{equation}
\label{SE}
i\hbar\frac{\partial \psi}{\partial t}=H\psi
\end{equation}
\noindent where  $i$ represent the imaginary unit, $\hbar$ is the Planck constant and $H$ is the Hamiltonian operator, which gives the total energy of the system under consideration being defined as the sum of kinetic and potential energy. This equation is unitary and linear and provides the fundamental law of motion of quantum systems. Nonetheless, it is useful to stress that the physics of RQM is best expressed in terms of the Heisenberg Equation (HE), since according to Rovelli's account only physical quantities evolve in time and not quantum states:
\begin{equation}
\label{HE}
\frac{d}{dt}A(t)=\frac{i}{\hbar}[H, A(t)]+\bigg(\frac{\partial A}{\partial t}\bigg)_H
\end{equation}
\noindent where $A$ is an observable and $[H, A(t)]$ is the commutator of the two operators $H$ and $A$;
\item \textbf{The projection postulate (Collapse rule):} Performing a measurement of a certain observable quantity $A$ on a given system $s$, relative to an observer $O$, the measurement interaction randomly projects the system in one of the possible eigenstates of the corresponding operator $A$. The probabilities for this stochastic transition are given by  the Born's rule;
\item \textbf{The Born's Rule:} If an observable quantity represented by the self-adjoint operator $A$ is measured:
\begin{itemize}
\item the result will be one of the eigenvalues $a_i$ of $A$;
\item the probability $P$ to obtain an eigenvalues $a_i$ is given by $\langle\psi|P_i|\psi\rangle$.
\end{itemize} 
\end{itemize}

The substantial difference introduced by RQM with respect to QM is that distinct observers  provide in general different descriptions of sequences of physical events, and such descriptions are equally correct. This is what Rovelli calls the ``Main Observation'' of relational quantum mechanics, and it is deduced by the ``third person problem'' which we are going to present. 

Let us consider the following idealized experimental situation: an observer $O$ is going to perform a measurement of a certain quantity $A$ on a system $s$. Suppose for the sake of the discussion that this magnitude can take just two (eigen)values $1, 2$, where $|1\rangle$ and $|2\rangle$ are the corresponding eigenstates. Before the measurement, say at time $t_1$, the system $s$ is in a superposition of the possible eigenstates of $A$, $a|1\rangle + b|2\rangle$ (where $|a|^2, |b|^2$ give the probabilities to find $s$ in $|1\rangle$ and $|2\rangle$ respectively). Assume also that performing the measurement at a later time $t_2$, $O$ will find the result ``1'', meaning that $s$ is projected into the state $|1\rangle$. 
Then, the sequence $\textbf{E}$ of physical events taking place in the lab can be schematically summarized as follows:
\begin{equation}
\left.\begin{aligned}
\label{E}
t_1 \longrightarrow t_2\\
a|1\rangle+b|2\rangle \longrightarrow |1\rangle
\end{aligned}\right\} = \textbf{E}
\end{equation}

\noindent This is just the usual quantum mechanical description of the measurement process, where the interaction between system and measuring device causes the suppression of the Schr\"odinger evolution and projects the quantum state onto one of the possible eigenstates of the observed quantity. Now, let us consider the perspective of another observer $P$ that (i) knows the initial states of $s$ and $O$, and (ii) describes quantum mechanically the interaction between $s$ and $O$. In this scenario the observer $P$ does not perform herself any measure on the complex system $s+O$ in the time interval $t_1-t_2$. Prior the measurement of $A$ on $s$ by $O$, $P$ knows that $O$ is in a neutral state not pointing at any particular result (i.e.\ neither in the ``1'', nor in the ``2'' direction), it is then in its ``ready state''. Such a ``ready'' state is correlated with the system $s$, which before the measurement is in a superposition of $|1\rangle$ and $|2\rangle$ states, as stated a few lines above. Given the linear dynamics of QM provided by \eqref{SE}, $P$ obtains the following correlation:
\begin{equation}
\left.\begin{aligned}
\label{Estar}
t_1 \longrightarrow t_2\\
(a|1\rangle+b|2\rangle)\otimes |O-ready\rangle \longrightarrow a|1\rangle\otimes|O_1\rangle + b|2\rangle\otimes |O_2\rangle
\end{aligned}\right\} = \textbf{E'}
\end{equation}

\noindent Thus, in the context of RQM the same sequence of physical events is described differently by two distinct observers; however, in virtue of the principle of relativity of states assumed by Rovelli, \emph{both} descriptions \textbf{E} and \textbf{E'} are equally correct and legitimate. According to $O$ the system $s$ has a sharp value for $A$ after the interaction, whereas for $P$ the system $s$ is not in the state $|1\rangle$ and the measuring device does not indicate any sharp value, meaning that the complex system $s+O$ is in a superposition of states represented by \eqref{Estar}. This is the specific and characteristic trait of RQM. 

This conclusion entails several consequences.\ Firstly, the separation between observed system and observer cannot be univocally determined, meaning that every system can play the role of observed system and of observer. In the third person scenario, not only $O$ observes the system $s$, but also it is part of the composite observed system $s+O$ according to $P$'s perspective. Secondly, as in special theory of relativity there is no privileged observer, meaning that all systems are equivalent: ``[n]othing a priori distinguishes macroscopic systems from quantum systems. If the observer $O$ can give a quantum description of the system $s$, then it is also legitimate for an observer $P$ to give a quantum description of the system formed by the observer $O$'' (\cite{Rovelli:1996}, p.\ 1644).\footnote{Referring to this, in \cite{Smerlak:2007} it is stated that ``[a]n observer, in the sense used here, does not need to be, say ``complex'', or even less so ``conscious''. An atom interacting with another atom can be considered an observer. Obviously this does not mean that one atom must be capable of storing the information about the other atom, and consciously computing the outcome of its future interaction with it; the point is simply that the history of its past interaction is is principle sufficient information for this computation'', p.\ 430, footnote 9.} Thirdly, in RQM different observers may provide distinct descriptions of the same sequence of events (cf.\ the Main Observation above). What is crucial to underline is that two different observers do not assign different \emph{probabilities} for possible measurement results, but rather they provide descriptions of different \emph{state of affairs}, as in the example discussed above. Consequently, RQM entails that the notion of state of a quantum system is relative to some observer. Thus, the notions of ``system'', ``measurement outcome'' and ``value of a variable'' are relational concepts.  

Furthermore, according to Rovelli's theory, QM provides a \emph{complete} description of the world: there is no deeper theory describing how absolute reality behaves. An interesting question to ask, then, is how different perspectives may coexist without generating contradictions. This query can be answered considering the third person problem discussed above. Taking into account the sequence \eqref{E} we know that $O$ performed a measurement of $A$ on the system $s$ obtaining a definite result. On the other hand, from \eqref{Estar} we infer that $P$ knows that $O$ performed a measurement of $A$ on $s$, obtaining information about the value of $A$. However, it should be emphasized that $P$ does not have any information about the measurement result obtained by $O$, i.e.\ $P$ does \emph{not} know which one among the possible alternatives has been actualized.\ $P$ does not have this information since it did not interact directly with $O$.\ What $P$ may predict, given its amount of knowledge and the quantum formalism, is just that $O$ made a measurement on $s$; however, it cannot predict the exact value that has been obtained.\ Thus, there is no contradiction between these two different perspectives. Clearly, $P$ can know $O$'s result through physical interaction: 
\begin{quote}
if $P$ knows that $O$ has measured $A$ (notation adapted), and then she measures $A$, and then she measures what $O$ has obtained in measuring $A$, consistency requires that the results obtained by $P$ about the variable $A$ and the pointer are correlated'' (\cite{Rovelli:1996}, p.\ 1652). 
\end{quote}

\noindent We are arrived to another tenet of RQM: information can be obtained only via physical interaction; what is important to underline for the purposes of the present essay is that in RQM information is concerned with the attribution of values to physical quantities.\footnote{For a technical discussion about the notion of information in relational quantum mechanics the reader may refer to \cite{Rovelli:1996}, Sections 2.5 and 3.} In sum, one can summarize the physical content of RQM as follows: 
\begin{quote}
\emph{quantum mechanics  provides a description of the world in terms of properties of physical systems relative to other systems functioning as observers, and such a description is complete}.
\end{quote}

Finally, to complete our qualitative survey of relational quantum mechanics, let us speak about the ontology of this theory. RQM employs a discrete ontology of events (or physical facts) in spacetime, recalling the Wittgensteinian idea according to which the world is the totality of facts, and not of things. More precisely, the ontology of the theory is given in terms of physical variables (as in classical mechanics) whose value change in interactions. The main difference between RQM and classical physics is that in the former theory observables take definite values only in interactions---taking into account also the contextual nature of quantum observables (cf.\ \cite{Kochen:1967}) and the limit imposed by Heisenberg uncertainty relations---and such values are relative to a particular observer. Against this background, it is worth noting that events are just observer-dependent change of physical values taking place in interactions among systems. According to RQM, thus, the world is just ``an evolving network of sparse relative events, described by punctual relative values of physical variables'' (\cite{Laudisa:2019b}). Moreover, variables in RQM can take values at certain time, and have no sharp values at other times, a fact referred as to the discreteness of quantum observables in Laudisa and Rovelli's introduction to RQM. 

Contrary to standard QM, relational quantum mechanics should not be considered as a theory about the behaviour of the wave function in spacetime.\ Remarkably, $\psi$ is just a convenient formal tool, useful for the computation of future quantum probabilities given a certain amount of knowledge at disposal of a particular observer. Thus, in RQM $\psi$ is not interpreted realistically, meaning that the wave function does neither represent a property of an individual system, nor a concrete object in physical space.
Contrary to the Schr\"odinger's picture where the wave function describes completely the state of a system and evolves unitarily in space and time according to \eqref{SE}, in the Heisenberg picture $\psi$ just encodes information about past interactions, and it changes only as a result of another interaction. Moreover, RQM provides a description of reality in terms of discrete events which are modeled as discrete changes in the relative state, ``when the information is updated and nothing else. What evolves with time are the operators, whose expectation values code the time-dependent probabilities that can be computed on the basis of the past quantum events'' (\cite{Smerlak:2007}, p.\ 431). This is why it has been claimed that the physical content of RQM is best described using Heisenberg equation \eqref{HE}.\footnote{Remarkably, the collapse postulate assumes different meanings in RQM and QM; since the wave function is not considered a physical object in the former, literally speaking nothing physical collapses in measurements interactions. Moreover, given that the proper dynamical description of RQM should be given in terms of the Heisenberg equation \eqref{HE}, it follows that the collapse of the wave function should be interpreted as an information update relative to a certain observer and concerning the value of some magnitude measured on a particular system.}

Hence, one can affirm that in RQM $\psi$ encodes the information referring to the values of physical magnitudes of a certain system relative to another system functioning as observer\footnote{As a consequence, in RQM also the notion of ``wave function of the universe'' present in several interpretations of quantum theory---as for instance in Everett's relative state formulation of QM, the Many Worlds interpretation, Bohmian mechanics, \emph{etc.}---is rejected.}:
\begin{quote}
[t]he state $\psi$ that we associate with a system $S$ is therefore, first of all, just a coding of the outcome of these previous interactions with $S$. Since these are actual only with respect to $A$, the state $\psi$ is only relative to $A$: $\psi$ \emph{is the coding of the information that $A$ has about $S$}. Because of this irreducible epistemic character, $\psi$ is but a relative state, which cannot be taken to be an objective property of the single system $S$, independent from $A$. Every state of quantum theory is a relative state (\cite{Smerlak:2007}, p.\ 431).
\end{quote} 

In sum, we can conclude this section by saying that ``the ontology of RQM is a sparse (``flash'') ontology of relational quantum events, taken as primitive, and not derived from any ``underlying'' representation'' (\cite{Laudisa:2019b}).

\section{Mereological Bundle Theory}
\label{bundle}

\subsection{Why Bundle Theory?}

At the end of the previous section it has been stated that relational quantum mechanics implements an event ontology.\ Nonetheless, in the literature concerning RQM it is not precisely stated what interacting physical systems are supposed to be.\ Referring to this, in many places it is even claimed that RQM rejects an object-oriented ontology, favoring an ontology of processes and relations (cf.\ for instance \cite{Candiotto:2017} and references therein). Contrary to this thesis, in the opinion of the present author Rovelli's theory can be made compatible with an ontology of properties from which objects can be easily defined. Indeed, here I propose a new metaphysical definition of physical systems in the context of relational quantum mechanics using the conceptual tools of mereological bundle theory, building on previous work contained in \cite{Paul:2017}. By shedding light on what kind of physical objects populate spacetime in RQM, I aim at clarifying the ontological picture of reality provided by this theoretical framework. 

In a nutshell, it is possible to summarize the central idea of the present essay by saying that physical systems in RQM should be defined as mereological bundles of observer-dependent properties varying in virtue of interactions. Given the central role played by quantum observables in RQM, it is more than plausible to claim that this framework is compatible with a form of realism towards \emph{properties} rather than substances.\ This proposal finds evidence and justification in Rovelli's work, since he stressed in several places the ontological priority of properties over states, and that RQM should be framed in the Heisenberg picture and not in Schr\"odinger's picture, as recalled in the previous section. Moreover, in order to provide RQM with an property-oriented ontology, it is crucial to note that Rovelli's theory discards the Aristotelian conception of object---a tradition centered on the notion of substance carrying attributes which dominated Western philosophy---in favor of an ontology of properties firstly proposed in David Hume's \emph{Treatise of Human Nature} (\cite{Hume:2007}). In this work Hume advances a devastating criticism of the notion of substance, more specifically of \emph{bare particulars}, i.e.\ propertyless substances that are things in themselves and property bearers. Referring to this, the Scottish philosopher writes that:
\begin{quote}
I wou'd fain ask those philosophers, who found so much of their reasonings on the distinction of substance and accident, and imagine we have clear ideas of each, whether the idea of \emph{substance} be deriv'd from the impressions of sensation or of reflection? If it be convey'd to us by our senses, I ask, which of them; and after what manner? If it be perceiv'd by the eyes, it must be a colour; if by the ears, a sound; if by the palate, a taste; and so of the other senses. But I believe none will assert, that substance is either a colour, or sound, or a taste. The idea of substance must therefore be deriv'd from an impression of reflection, if it really exist. But the impressions of reflection resolve themselves into our passions and emotions; none of which can possibly represent a substance. We have therefore no idea of substance, distinct from that of a collection of particular qualities, nor have we any other meaning when we either talk or reason concerning it. The idea of a substance as well as that of a mode, is nothing but a collection of simple ideas, that are united by the imagination, and have a particular name assign'd them, by which we are able to recal, either to ourselves or others, that collection (\cite{Hume:2007}, p.\ 16).
\end{quote}

In this passage Hume seeks to show that philosophers do not have any definition of what a substance really is: neither it can be defined from senses and perceptions, nor from abstract reasoning. In particular, he stresses that we cannot think an object without attributing properties to it. Thus, the notion of bare particular would not be actually conceivable. More precisely, Hume claims that our beliefs about substances derive from an illusion of thought: we perceive that objects change in time and modify (often radically) their attributes; then, we are led to ascribe a temporal identity to those objects even though we assign them a set of different and contradicting properties in time. Thus, ``[i]n order to reconcile which contradictions'' Hume says, ``the imagination is apt to feign something unknown and invisible, which it supposes to continue the same under all these variations; and this unintelligible something it calls a \emph{substance}, or \emph{original and first matter}'' (\cite{Hume:2007}, p.\ 146). As \cite{Robinson:2018} notes, Hume's critique of the notion of substance is very similar to his criticism of the concept of causation: it is a projection and a tendency of our mind, that seeks to associate the things that we perceive with the passage of time. If bare particulars do not exist, what are then the objects populating space and time? Hume answered this question resorting to what is now called bundle theory, according to which objects are defined as bundles---i.e.\ collections---of properties. Individual objects, therefore, are simply a collection of properties colocated in spacetime, without any bare substratum carrying them.\footnote{Various bundle theories propose different criteria to individuate objects and distinguish them with respect to mere collections of properties.\ For space reasons I will not enter here in these details. Nonetheless, I will explicitly address how objects are individuated according to mereological bundle theory, which will be employed to define what physical systems are in RQM.} For instance, in this account a billiard ball is simply the collection of its attributes, e.g.\ its roundness, its weight, its color, \emph{etc.}, however, there is no bare substance of a propertyless billiard ball bearing such properties. In the XX century Hume's proposal evolved into various bundle theories which have been endorsed and improved by several prominent philosophers.\footnote{The interested reader may refer for instance to \cite{Russell:1940} \cite{Simons:1994}, \cite{Williams:1953}.}\ Moreover, bundle theories have been introduced recently also in the philosophy of physics in order to provide quantum mechanics and quantum field theory with a clear ontology of properties (cf.\ \cite{Lombardi:2013}, \cite{Lombardi:2016} and \cite{Kuhlmann:2010aa} respectively).\footnote{For a similar proposal see \cite{Falkenburg:2007}, where she argues in favor of an ontology of properties.}
To this regard, another motivation to discard the notion of object as substance is given by the quantum mechanical formalism itself, which does not allow to conceive quantum objects as independent, localized entities along the lines of classical mechanics. Indeed, QM entails essential features as indistinguishability, contextuality and non-separability, which make quantum particles inherently different from their classical counterparts. Thus, the notion of individual system should strongly differ in the quantum and classical regime. For all these reasons, taking into account a different philosophical tradition, where properties (without substances) are the fundamental elements of the ontology, can be helpful in order to provide a novel definition of what objects are in the context of quantum theory.\footnote{To this regard cf.\ \cite{Lombardi:2016}, p.\ 127.}  
  
Following these latter examples, I am going to introduce mereological bundle theory, a metaphysical framework which will give us the necessary tools to propose a meaningful notion of object suitable for relational quantum mechanics. 

\subsection{Abolishing Substances: Mereological Bundle Theory}

MBT is a metaphysical theory proposed by L.A. Paul (cf.\ \cite{Paul:2017}) providing a one-category ontology of the world in terms of properties\footnote{Here I will refer to what Paul calls the ``mosaic model'' of MBT.}; such an ontology, moreover, employs a unique world-making relation, i.e.\ mereological composition. This proposal is particularly apt for the aim of the present essay, since in Paul's view spacetime, matter and complex physical objects are ``constructed from mereological fusions of qualities, so the world is simply a vast mixture of qualities, including polyadic properties (i.e., relations)'' (\emph{ibid.}, p.\ 33). Hence, mereological bundle theory abolishes the traditional distinction between an object and its properties, given that the former notion is ontologically reduced to the latter. On the other hand, this view preserves the possibility for objects---properly understood as mereological fusion of properties---to be located in spacetime, since they are ``qualitative fusions that are fused with spatiotemporal relations or relational properties'' (p.\ 34).\footnote{In traditional bundle theories such as those defended by Russell, Simons and Williams we find a distinction between universal properties and particular objects, therefore, these should not be considered as a one category ontology, contrary to MBT. For details see \cite{Paul:2017}, p.\ 36. To this regard, the distinction between universal and particulars seems to be present also in \cite{Lombardi:2013} and \cite{Lombardi:2016}. Indeed, in these works self-adjoint operators represent physical observables, which in turn correspond to ``type-properties''. On the other hand, eigenvalues of self-adjoint operators correspond to the values of observables, and to ``case-properties'' from an ontological perspective. Here type-properties should be conceived as universal attributes which can have countless particular instances of case-properties. However, \cite{Lombardi:2016} claim that their approach remains neutral with respect to the existence of universal properties. It is worth noting that in mereological bundle theory non-instantiated universals do not exist, so that its ontology is not inflated with uninstantiated properties.} Thus, also spacetime is ontologically reduced to the spatial relations in which objects stand with respect to each other; MBT, therefore, endorses a relational view of spacetime.\ It is straightforward to note that this metaphysical view shares fundamental features with RQM since both theories (i) eliminate the notion of object as propertyless substance in favor of an ontology of properties, and (ii) do not consider spacetime to be a substance \emph{per se} over and above relational (i.e.\ spatial) properties. Hence, MBT seems to be a perfect candidate in order to provide a correct definition of what physical objects are in the context of Rovelli's quantum theory.

In a nutshell, conforming to MBT the elementary, fundamental building blocks of our world are properties (including relations), and everything else is mereologically composed from them. In this metaphysical theory properties are literally taken as objects (and part of objects) bundled together via the composition relation.\footnote{The basic axioms of MBT are given in \cite{Paul:2017}, p.\ 38.\ In what follows I am not going to introduce them since they are not essential for the purpose of the present essay. Nonetheless, one can say that the primitive mereological notion at play in MBT is that of ``proper part'', which is irreflexive, asymmetric and transitive. Such a concept is complemented by the other notions of parthood and composition, which are essential in the context of classical extensional mereology.}\ Furthermore, properties---and hence objects---are individuated since they are fused to spatiotemporal relations; therefore, MBT avoids well-known issues related with the notions of co-location and compresence typical of traditional bundle theories. Indeed, in virtue of MBT's view about spacetime, spatiotemporal locations are defined in Paul's account as $n$-adic properties which individuate objects and assign them a relational identity. 

To this regard, it is worth noting that mere collections of properties do not form an object in Paul's theory: properties must be bundled together using exclusively mereological composition which include also spatiotemporal relations, so that strange sums of properties like ``being squared'' and ``being triangular'' cannot count as a proper object in this framework. Remarkably, according to Paul's bundle theory the derivative ontological structure of the macroscopic world is reduced to the fundamental entities appearing in the vocabulary of our most advanced physical theories, such as quantum mechanics and quantum field theory.\ These fundamental entities, in turn, are constituted by inherent and relational properties that form their mereological structure. What is metaphysically interesting is that, conforming to MBT, sums of properties do not generate new ontological categories: the emergence of complex objects from the Plank to the macroscopic and cosmological scales does not entail novel ontological categories with respect to properties. 

At this point, however, an objection can be raised: properties seem to be essentially different from the discrete, concrete material objects populating and interacting in spacetime, so that we should think of them as a different and separate category of being. To this worry, Paul replies that 
\begin{quote}
[i]t is just a mistake to think properties cannot be chunky, concrete, complete, or independent. They are chunky, and concrete, and complete, and independent---because some of them are chunky, concrete, complete, and independent. In particular, some of the properties that are ordinary objects are chunky, concrete, complete, and independent. (As are some of the fundamental physical properties, such as field intensities. [...]) Do not be tempted by the fallacious idea that fusing is what somehow ``makes'' the ordinary object (which is a fusion of properties) chunky or substantial. That's not how fusing works: it makes many into one, it doesn't make non-substances into substances or abstract things into concrete ones (\cite{Paul:2017}, p.\ 42). 
\end{quote}

\noindent Referring to this, the mosaic model of mereological bundle theory claims that extended objects are composed by two sorts of properties, qualitative and spatiotemporal, where the latter should be thought as relational properties, as repeatedly underlined above; ``Rocks, persons, stars, and abstract objects are all fusions built from quality fusions then fused together by spatiotemporal composition. Such fusions, in addition to being complex constructions of quality and spatiotemporal fusions, are also plain vanilla property fusions, where the properties fused are the whole (distributed) properties of the object'' (\emph{ibid.}, p. 45).

Another interesting feature of mereological bundle theory concerns its treatment of identical objects, a topic which is essential for the present discussion because our aim is to apply MBT to quantum objects. What is important for our purpose is to allow that objects with the same properties exist and are also numerically distinct. For instance, we want to be able to say that all electrons are characterized by the same properties of mass, charge and spin, and that they are numerically distinct. Since MBT cannot rely on haecceities or substances in order to individuate objects and to distinguish between identical instances of the same object, it has to resort to another strategy, i.e.\ to admit as a primitive fact the existence of distinct objects which are fusions of the same properties. In this way, one has qualitatively indistinguishable, but numerically distinct objects. The problem of individuating indistinguishable but distinct objects stems from Leibniz's Principle of the Identity of Indiscernibles (PII), which affirms that if two objects $a,b$ share the same properties, then they are identical. This issue is particularly delicate and deserves some attention. It is well-known that PII has been given many interpretations, but Paul considers the following two:
\begin{enumerate}
\item $a$ and $b$ share all their properties, including those such as ``being identical to $a$'';
\item  $a$ and $b$ share all their pure intrinsic and extrinsic properties.
\end{enumerate}

\noindent For our discussion it is crucial to underline that pure properties exclude primitive identity-based properties as for instance ``being this laptop'', ``being Olga'', ``being that car'', \emph{etc.}. Now, if PII is interpreted according to (1), then every philosopher agrees that it is true; however, PII is not necessarily true if indiscernibility is interpreted in the sense of (2). The usual problem for bundle theorists, Paul argues, is that bundle theory is generally considered as entailing the truth of PII according to interpretation (2). This entailment, she adds, is due to another implicit premise, the ``Supervenience of Identity Thesis'', which states that the property ``being identical to $a$'' reductively supervenes on $a$'s pure intrinsic and extrinsic properties. Nonetheless, we stated a few lines above that pure properties do not concern identity-based properties, thus, the mereological bundle theorist should reject the Supervenience of Identity Thesis, exactly as the substance theorists do. However, if the latter appeals to the intrinsic essence of objects, the former needs to say that identity facts just ``supervene on the object themselves: i.e., the identity of $a$ supervenes on $a$, and that's all'' (\cite{Paul:2017}, p.\ 50). This rejection leads to a new primitive assumption for MBT, the ``ungrounded difference''. This primitive fact allows us to claim that, according to the mosaic model of mereological bundle theory, the fusions of identical properties generate barely different objects; thus, the mereological bundle theorist can accommodate the existence of identical, but numerically distinct objects.\footnote{To this regard, Paul discusses in depth how MBT deals with identical objects located at the very same spatial location considering the case of $n$-bosons localized in the very spacetime point. For spatial reasons we will not discuss this issue in the present essay.}

Finally, to conclude our presentation of MBT and to make it compatible with quantum theory, we have to take into account typical features of quantum particles.\footnote{Here I am slightly modifying MBT by applying the quantum formalism to it.} It is well-known that in quantum mechanics the word ``particle'' does not refer to point-like material objects with well-defined properties as in classical mechanics. This must be also valid in the context of MBT: here different species of quantum particles refer to fusions of specific properties, as for instance having a certain mass, a given charge, a particular spin \emph{etc.}. Moreover, in order to be empirically adequate, this ontology of properties must take into account the contextual nature of quantum theory, following the work of \cite{Lombardi:2013}, and \cite{Lombardi:2016}. These authors recall that in virtue of the quantum formalism not every property constituting a quantum system can have well-defined value. More precisely, the Kochen-Specker theorem entails that it is not possible to ascribe simultaneously well-defined values to all the observables attributable to a quantum system without generating contradictions. 
Thus, also in MBT not every property defining a given system will have an actual, definite value. For instance, in virtue of the quantum formalism, a certain particle will not have simultaneously well-defined position and momentum, although the properties of ``having position'' and  ``having momentum'' can be ascribed to quantum objects. Consequently, the fundamental building blocks of our reality are defined as bundles of fused properties, where some of them remain metaphysically indeterminate in virtue of the contextual nature of quantum formalism (cf.\ \cite{Calosi:2020}). For instance, an electron is defined by the fusion of its intrinsic properties like mass, charge, being spin$-1/2$, and extrinsic properties as momentum, velocity, energy, position \emph{etc.}. Remarkably, whereas the former class of properties have the same values for every instance of electron (i.e.\ every electron will have the same mass, charge and will be a spin$-1/2$ particle), the latter (among which there are also relational properties) depend on the specific interaction of the particular particle under consideration. In any case, the fusion of these attributes is what defines each particular electron according to mereological bundle theory. This examples can be extended to every other family of particles. Composite systems, like nucleons, atoms and molecules are mereologically---and hence, ontologically---dependent on the composition of different species of particles. Let us now make the last step, and apply MBT to relational quantum mechanics.

\section{A New Definition of Physical Objects in RQM} 
\label{RQMBTA}

\subsection{The Mereological Bundle Theory Approach to RQM}

Having introduced the essential ideas of mereological bundle theory, in this section I will propose a new metaphysical interpretation of physical systems in relational quantum mechanics in terms of Paul's theory.

According to RQM a physical system can be characterized ``by a family of yes/no questions that can be meaningfully asked to it'' (\cite{Rovelli:1996}, p.\ 1655), where such questions are measurements\footnote{From von Neumann's treatise on quantum mechanics (\cite{vonNeumann:1955}), it became popular to define measurements as question asked on quantum systems. This is still a standard practice in the field of quantum logic and quantum information.} that can be performed on physical observables, i.e.\ properties, attributable to the system under consideration. An observer $O$ may ask a set of potentially \emph{infinite} questions $Q_1, Q_2, \dots, Q_n$ to the system $s$, obtaining the string 
\begin{align}
\label{string}
(e_1, e_2, \dots, e_n)
\end{align}

\noindent where each $e_i$ represent a specific answer.\ Nonetheless, it is worth stressing that RQM postulates that ``there is maximum amount of \emph{relevant information} that can be extracted from a system'' (First postulate of RQM \emph{ibid.}, p.\ 1657). From this principle, it follows that a complete description of a physical system $s$ is given in terms of the string
\begin{align}
\label{string2}
[e_1, e_2, \dots, e_k]
\end{align}

\noindent with $k<n$ which is a subset of \eqref{string}. Given that in RQM information about systems is obtainable only through interaction, and since questions are essentially measurements on the system $s$ performed by some observer $O$, it follows that \eqref{string2} contains the information that $O$ has (extracted) about the system $s$. It is worth noting that \eqref{string2} provides the \emph{relevant} information about $s$, where ``[t]he relevant information is the subset of \eqref{string} [equation number adapted] obtained by discarding the $e_i$ that do not affect the outcomes of future questions'' (\emph{ibid.}, p.\ 1656).\footnote{It is important to underline that repeating a measurement of an observable on a given system and obtaining the same result will not increase the information about the system at hand.} In sum, the string \eqref{string2} is the knowledge that an observer $O$ has about the system $s$, and the subscript $k$ refers to the number of questions asked to $s$ and that characterize it.\ Clearly, this string represents the description of $s$ relative to $O$; indeed, another observer $P$ may ascribe to $s$ a different list of properties, as we have seen in the third person scenario described in Section \ref{RQM}. Referring to this, it is important to underline that, contrary to the case of classical mechanics, the amount of information that an observer can extract from a quantum system is finite in virtue of the algebraic structure of quantum observables. Although the information about a particular system is finite, RQM postulates that it is always possible to acquire new information about it (Second postulate of RQM). These two claims are not in tension, since the second postulate merely refers to the fact that, even if one knows the ``state'' of a certain system, it is possible to gain new information performing a measurement of a given observable $A$, such that the system $s$ at hand is not in an eigenstate of $A$. Given that the amount of information has a upper bound as imposed by the first postulate,
\begin{quote}
when new information is acquired, part of the old relevant information must become irrelevant. In particular, if a new question $Q$ (not determined by the previous information gathered) is asked, then $O$ should lose (at least) one bit of the previous information. Thus, after asking the question $Q$, new information is available, but the total amount of information about the system does not exceed $N$ bits (\emph{ibid.}, p.\ 1658).
\end{quote}

\noindent Thus, in agreement with the Heisenberg picture and the algebraic approach to quantum mechanics, also in Rovelli's theory a system is defined via the family of observables that are attributed to it. For instance, meaningful questions concerning a quantum particle may refer to the possibility to find it in a certain spatial region, about its momentum or its spin in some direction, \emph{etc}..

As a consequence, since in relational quantum mechanics physical systems are defined via a specific set of observables, and the latter are usually associated to their instantiated properties, it is then plausible to say that such systems may be characterized as mereological bundles of properties, as anticipated in Section 3.1. In this manner RQM can be provided with an property-oriented ontology where objects are straightforwardly defined, because in MBT properties do correspond to objects, as already stated. However, given that the values of the properties of quantum systems in relational quantum theory are observer-dependent, we define the objects of this theoretical framework as mereological bundles of properties whose values depend on the perspective from which they are observed. Referring to this, one should say more precisely that there is a set of inherent properties characterizing a certain species of particles---such as mass, charge, spin \emph{etc.}---which are not observer-dependent, so that their value remains constant.\footnote{If intrinsic properties of quantum particles could change, then different observers could observe particles with diverse identities. This fact, however, would entail disastrous empirical consequences to be admitted.} On the contrary, the values of extrinsic properties (as for instance energy, position, momentum, \emph{etc.}) are observer-dependent and change relatively to specific observers. Furthermore, since these bundles of properties are subjected to the formalism of QM, it follows that not all observables associated with a certain system, that is, not all properties constituting the system, 
can have definite values; this fact must be respected to avoid contradictions with the Kochen-Specker theorem and the contextuality of quantum theory. 

In sum, this proposal suggests that physical systems in RQM should be defined as mereological bundles of properties, where (i) the intrinsic properties characterizing a certain species of particles have constant values, (ii) the extrinsic qualities take definite values relative to particular observers, and (iii) not every observable defining a system can have a definite value in virtue of the contextual nature of the quantum formalism. To give an example, in RQM an electron is composed by its inherent properties characterizing this species of particles as for instance its mass, charge, having spin$-1/2$, \emph{etc.}, and by its extrinsic properties as momentum, energy, angular momentum, or position. These latter have relational and contextual values which depend on the specific interactions of the particle under consideration with different observers; not all these extrinsic properties have definite values, in agreement with the Kochen-Specker theorem. Clearly, all electrons are composed of the same intrinsic and extrinsic properties fused together, where the former have the constant value for every instance of single electron, and the latter depend on the specific interactions of individual electrons. Since MBT admit the existence of numerically distinct objects which are bundles of the same properties, we can meaningfully speak of the class, or the family of electrons. This example can be easily extended to every family of quantum particles. 

A remarkable consequence of this proposal is that, under this reading, RQM is committed to a realism towards properties; thus, from the perspective of mereological bundle theory, one can properly speak about material objects in motion in spacetime, whose attributes vary in relation to different observers. Hence, not only one can define what physical systems are according to relational quantum mechanics, but also it is possible to clarify its event ontology. Indeed, with this new approach to RQM we can properly claim that there are interactions among physical objects in spacetime. The notable feature of this proposal is that RQM can speak about interactions without resorting to the notion of bare substance, which is abolished in this theoretical framework.


\subsection{Relational QM and Structural Realism}
\label{MSR}

In a recent and interesting paper Laura Candiotto argues that relational quantum mechanics is an instantiation of the ontology proposed by Ontic Structural Realism (OSR), according to which the fundamental building blocks of nature are not objects, but relations (cf.\ \cite{Candiotto:2017}). More precisely, Candiotto claims that, according to RQM, relations constitute the fundamental elements of our reality, i.e.\ relations are the only primitives in this interpretation of QM. Furthermore, given the event ontology of Rovelli's theory, Candiotto says that in interactions physical systems change their properties, and the net of such physical interactions constitute our world. Hence, she concludes that in RQM only relations are fundamental, relata are not. Quoting Rovelli himself, Candiotto affirms that in RQM there are no objects entering into a given relation: it is the relation which generates objects. From this viewpoint, given that in RQM (i) there is an ontological priority of relations over relata, and (ii) that relations form real structures, Candiotto claims that the natural framework to cast relational quantum mechanics and to understand its metaphysical content is OSR, a well-known philosophical perspective which argues in favor of an ontology of pure structures.\footnote{Cf.\ \cite{Ladyman:2007b} and \cite{Ladyman:2014} for details on this perspective.}

Although I agree with Candiotto's primary aim, i.e.\ to argue that RQM should be considered a realist quantum theory, I disagree with the claim that relational quantum mechanics denies the notion of object \emph{altogether}. In the opinion of the present author Candiotto is absolutely right in pointing out that ``RQM implies a critique of the notion of object.\ The objects denied by the RQM are the objects' things which characterize naive realism; denying their existence does not imply that there is nothing or nothing is real. The challenge is to think reality in relational terms, engendering thus a new way of understanding ``objects' '' (\cite{Candiotto:2017}, p.\ 6). 
Indeed, as it has been argued in the previous sections, Rovelli's theory demands a notion of object that is completely detached from the classical concept of bare substance carrying attributes. Furthermore, denying the existence of such substances in relational quantum mechanics does not entail that Rovelli's theory discards \emph{tout court} the notion of object, as has been shown taking into account the ontology of properties shaped applying MBT to RQM. Thus, a property-oriented ontology can be implemented in this framework, so that the notion of object can be retained. As a consequence, then, there is argumentative room to claim that relations are fundamental in RQM, \emph{as well as} objects (if this notion is properly understood). This conclusion, in turn, entails that RQM would be compatible with a Moderate form of Structural Realism (MSR) as presented in \cite{Esfeld:2011aa}, where not only relations are fundamental and primitive notions of the theory, but also \emph{relata}, i.e.\ concrete physical objects---in this case mereological bundles of properties---standing in such relations are so.  

More precisely, according to MSR there is no ontological priority between relations and relata: both are considered equally fundamental. Interestingly, supporters of this form of structural realism claim that the distinction between objects and relations---and properties more in general---is not ontological but only conceptual, in the sense that although such a distinction is present in our language, it does not reflect a real dichotomy of the world. Consequently, as \cite{Esfeld:2011aa} argue:
\begin{quote}
there is no point in enquiring into the relationship between objects and properties, including relations or structures, and, in particular, to talk in terms of a mutual ontological dependence between objects and properties, including relations or structures, or an ontological priority of the one over the others. There are not two types of entities, objects and properties including relations or structures, that entertain a certain relationship of ontological dependence. The dependence is only conceptual.
\end{quote} 

In sum, both relata and relations exist, and there is no question concerning the ontological priority of one category over the other. This feature of MSR reflects what has been proposed in the previous section, where it has been argued that systems in RQM are mereological fusion of properties and relations.\ Given that objects in this bundle theory \emph{are} properties and relations, it follows that in this new interpretation of RQM both relata and relations are equally fundamental. Hence, it follows that, in this new reading of Rovelli's theory a structure consists of a network of physical relations among objects interacting in spacetime. Thus, we conclude that since it is possible to provide relational quantum mechanics with a property-oriented ontology where object are easily defined, one should admit the fundamentality of these objects together with relations. 
\vspace{2mm}

In conclusion, from what has been argued so far, it is fair to claim that relational quantum mechanics should be included among the full-fledged realistic interpretations of quantum mechanics. Referring to this, it is worth noting that Rovelli's theory postulates the existence of mind-independent, concrete, physical objects interacting in spacetime. In this framework physical objects exist per se, there is no need of minds with some sort of creative power in the ontology of RQM. Moreover, this theory shows clearly the three-dimensional character of scientific realism, as \cite{Chakravartty:2017} Section 1.2 puts it. Indeed, (i) RQM endorses a form of metaphysical realism, i.e.\ the idea that the world (and the objects that compose it) exists independently of human minds perceiving or observing it, (ii) from a semantic perspective a supporter of RQM will believe in the statements, explanations and predictions of this theory, in the sense that relational quantum mechanics makes (approximately) true statements about the features of reality, finally (iii) the explanations of physical phenomena in RQM provide actual knowledge of the world to the supporter of the theory. Hence, the metaphysical, semantical and epistemological character of scientific realism are maintained in the context of Rovelli's theory.

Hence, since RQM can be provided with a property-oriented ontology which is compatible with moderate structural realism, and that the three fundamental aspects of scientific realism are respected, one can safely conclude that relational quantum mechanics must be included in the list of realist approaches to quantum theory.

\section{Conclusion}
\label{conc}

Elaborating on Paul's mereological bundle theory and following the proposals for a quantum ontology of properties advanced by \cite{Lombardi:2013} and \cite{Lombardi:2016}, in the present essay I proposed a new metaphysical interpretation of physical systems in the context of relational quantum mechanics, characterizing them as mereological bundles of properties. Essentially, this reading of RQM defines quantum systems as bundles of properties which can take contextual values with respect to particular observers. Such an ontology of properties is motivated by the principles of relational quantum mechanics, since in this theory observable quantities assume a primary ontological status, following the Heisenberg's picture and the algebraic approach to QM. Furthermore, this new interpretation elucidates the event ontology of Rovelli's theory, making it clear what kind of items populate the world in RQM. 
As a consequence of this proposal, RQM is made compatible with scientific realism, if the latter is intended to be about properties. More precisely, in this essay it has been claimed that RQM is compatible with a moderate form of structural realism, since also objects, i.e.\ the \emph{relata}, can be meaningfully defined in this theoretical framework. Thus, in the opinion of the present author, relational quantum mechanics should be incorporated among the full-fledged realistic interpretations of quantum theory.

Let me conclude by saying that although in this paper I tried to provide RQM with an object-oriented ontology, a lot of philosophical work remains to be done in order to clarify the rich metaphysical implications of this theory, as for instance what interpretation of probabilities is best suited for this framework, the problem of locality, the extension of this approach to the standard model of particle physics, \emph{etc.}. This is a positive fact, since both physicists and philosophers will have the opportunity to discuss the foundations of quantum theory from a new, hopefully fruitful perspective.
\vspace{5mm}

\noindent \textbf{Acknowledgements:} I warmly thank Carlo Rovelli, Olimpia Lombardi, Claudio Calosi and his group at the University of Geneva, Cristian Lopez and Olga Sarno for helpful and positive comments on previous drafts of this paper. This research is financially supported by the Swiss National Science Foundation (Grant No. 105212-175971).  

\clearpage

\bibliographystyle{apalike}
\bibliography{PhDthesis}
\end{document}